\RequirePackage{fix-cm}
\documentclass[smallextended]{svjour3}
\makeatletter\if@twocolumn\PassOptionsToPackage{switch}{lineno}\else\fi\makeatother

\usepackage{amsfonts,amssymb,amsbsy,tabulary,amsmath}
\smartqed  
\usepackage{graphicx}
\usepackage{url,multirow,morefloats,floatflt,cancel,tfrupee}
\makeatletter

\AtBeginDocument{\@ifpackageloaded{textcomp}{}{\usepackage{textcomp}}}
\makeatother
\usepackage{colortbl}
\usepackage{xcolor}
\usepackage{pifont}
\usepackage[nointegrals]{wasysym}
\makeatletter

\def\mcWidth#1{\csname TY@F#1\endcsname+\tabcolsep}

\def\cAlignHack{\rightskip\@flushglue\leftskip\@flushglue\parindent\z@\parfillskip\z@skip}
\def\rAlignHack{\rightskip\z@skip\leftskip\@flushglue \parindent\z@\parfillskip\z@skip}

\@ifundefined{etal}{}{}

\usepackage{ifxetex}
\ifxetex\else\if@twocolumn\@ifpackageloaded{stfloats}{}{\usepackage{dblfloatfix}}\fi\fi

\AtBeginDocument{
\expandafter\ifx\csname eqalign\endcsname\relax
\def\eqalign#1{\null\vcenter{\def\\{\cr}\openup\jot\m@th
  \ialign{\strut$\displaystyle{##}$\hfil&$\displaystyle{{}##}$\hfil
      \crcr#1\crcr}}\,}
\fi
}

\AtBeginDocument{%
  \@ifpackageloaded{endfloat}%
   {\renewcommand\efloat@iwrite[1]{\immediate\expandafter\protected@write\csname efloat@post#1\endcsname{}}}{\newif\ifefloat@tables}%
}%

\def\BreakURLText#1{\@tfor\brk@tempa:=#1\do{\brk@tempa\hskip0pt}}
\let\lt=<
\let\gt=>
\def\processVert{\ifmmode|\else\textbar\fi}

\@ifundefined{subparagraph}{
\def\subparagraph{\@startsection{paragraph}{5}{2\parindent}{0ex plus 0.1ex minus 0.1ex}%
{0ex}{\normalfont\small\itshape}}%
}{}

\newcommand\role[1]{\unskip}
\newcommand\aucollab[1]{\unskip}
  
\@ifundefined{tsGraphicsScaleX}{\gdef\tsGraphicsScaleX{1}}{}
\@ifundefined{tsGraphicsScaleY}{\gdef\tsGraphicsScaleY{.9}}{}
\def\checkGraphicsWidth{\ifdim\Gin@nat@width>\linewidth
	\tsGraphicsScaleX\linewidth\else\Gin@nat@width\fi}

\def\checkGraphicsHeight{\ifdim\Gin@nat@height>.9\textheight
	\tsGraphicsScaleY\textheight\else\Gin@nat@height\fi}

\def\fixFloatSize#1{}
\let\ts@includegraphics\includegraphics

\def\inlinegraphic[#1]#2{{\edef\@tempa{#1}\edef\baseline@shift{\ifx\@tempa\@empty0\else#1\fi}\edef\tempZ{\the\numexpr(\numexpr(\baseline@shift*\f@size/100))}\protect\raisebox{\tempZ pt}{\ts@includegraphics{#2}}}}

\AtBeginDocument{\def\includegraphics{\@ifnextchar[{\ts@includegraphics}{\ts@includegraphics[width=\checkGraphicsWidth,height=\checkGraphicsHeight,keepaspectratio]}}}

\DeclareMathAlphabet{\mathpzc}{OT1}{pzc}{m}{it}

\def\URL#1#2{\@ifundefined{href}{#2}{\href{#1}{#2}}}

\def\UrlOrds{\do\*\do\-\do\~\do\'\do\"\do\-}%
\g@addto@macro{\UrlBreaks}{\UrlOrds}

\edef\fntEncoding{\f@encoding}

\makeatother

\newif\ifmultipleabstract\multipleabstractfalse%
\usepackage[authoryear]{natbib}

\usepackage[T1]{fontenc}

\makeatletter	

\AtBeginDocument{%
  \@ifpackageloaded{endfloat}%
   {
\renewcommand*\efloat@process[2]{%
  \ef@ifct{#1}{%
    \expandafter\immediate\expandafter\closeout\csname efloat@post#1\endcsname
    \ef@setct{#1}{0}%
    \clearpage                                                         
        
    \efloat@ifflag{#2list}{
      {\normalsize\efloat@listof{#2}}
    }{}%

    \efloat@ifflag{#2head}{%
      \section*{\@nameuse{#2section}}
      \suppressfloats[t]
    }{}

    \markboth                                                          
      {\expandafter\uppercase\expandafter{\csname #2section\endcsname}}
      {\expandafter\uppercase\expandafter{\csname #2section\endcsname}}

    \def\efloat@type{#2}%
    \processdelayedfloat@hook
    \@nameuse{process#2s@hook}%
    \clearpage
    \@input{\jobname.#1}%
  }{}}
   
   }{}%
}%

  \makeatother

\begin{document}

\title{Optimized cutting off transit algorithm to study stellar rotation from PLATO mission light curves}


\titlerunning{Optimized Cutting off transit algorithm to study stellar rotation}     


\author{L. de Almeida \and F. Anthony  \and  A. C. Mattiuci \and M. Castro \and J. S. da Costa  \and R. Samadi \and J. D. do Nascimento Jr.}


\institute{Leandro de Almeida$^1$\\
        \email{dealmeida.l@fisica.ufrn.br}\\
         Francys Anthony$^1$, Ana Carolina Mattiuci$^1$,  Matthieu Castro$^1$, Jefferson Soares da Costa$^2$, 
         R\'eza Samadi$^3$, Jos\'e Dias do Nascimento J\'unior$^1$
        \\
        \\
        $^1$Departamento de F\'isica Te\'orica e Experimental, Universidade Federal do Rio Grande do Norte, Natal, RN, Brazil \\ 
        $^2$ Escola de Ci\^encias e Tecnologia, ECT, Universidade Federal do Rio Grande do Norte, Natal, RN, Brazil\\
        $^3$LESIA, Observatoire de Paris, Universit\'e PSL, CNRS, Sorbonne Universit\'e, Univ. Paris Diderot, Sorbonne Paris Cit\'e, France\\
}

\date{Received: date / Accepted: date}

\maketitle

\begin{abstract}
Measuring the stellar rotation of one of the components in eclipsing binaries (EBs) or planetary systems is a challenging task. The difficulty is mainly due to the complexity of analyzing, in the same light curve, the signal from the stellar rotation mixed with the transit signal of a stellar or sub-stellar companion, like a brown dwarf or planet. There are many methods to correct the long-term trend of the light curve. However, the correction often erases the signal of the stellar rotation from spots crossing the visible stellar disk and other weaker signals like planets. In this work, we present the DiffeRencial flUx Method of cuTting Off biNariES (DRUM TONES) to identify the signal of the binary transits and disentangle it from stellar rotation planet signals. We present our technique with applications to EBs from CoRoT, Kepler, Kepler K2 and TESS missions. We also applied our method to simulated synthetic EB from the PLATO mission. Our method shows good agreement in the determination of stellar rotation periods for few observed targets from last space missions, as well it is naturally useful for future European missions, such as PLAnetary Transits and Oscillations of stars (PLATO). 
\end{abstract}

\section{Introduction}

Disentangle the transit signal of a binary or a planet from the stellar signal, for example, is essential to the study of stellar rotation and to better understanding the stellar variability. In this context, especially the rotation of binaries has an impact in many branches of astronomy and, as described by  \citet{duch2013}, is a major point, because at least half of stars are binary systems. A number of studies involving eclipsing binaries (EBs) have been focused on tidal circularization (e.g. \citet{koch1981,duque1991,meibom2005,van2016}), and deal with the relations between the rotation period and orbital period. Most of the studies involving photometric observations of EBs allow a relative direct measurement of the system's orbital period, and also can be used to characterize the relative radius and luminosity ratios of the pair of stars. While the measurement of the orbital period of EB is relatively easy, the determination of the rotation period of one of the components is a challenge task. This is due to the difficulty to distinguish the \textbf{transit signals} and general trends of the light curve from the signal of spots modulation \textbf{from} the star. These general trends are the sum of instrumental artifacts and systematic trends that are characteristic of each instrument and can cause \textbf{false positive} modulations on the light curve, and an automatic pipeline can interpret this as planet transit or modulation due to spots. For cool stars with external convective zones and spots across the stellar disks, the analysis of the spot modulation present on the light curve can give us the possibility to measure the rotation period of the star \citep{basri2011, meibom2011, donascimento2014}. However, to access this information we carefully need to manipulate the light curves where all signals components are mixed.\\

Indeed, despite some advances, there are difficulties inherent to studying rotational modulation in eclipsing binaries and planetary systems. During the last 15  years, we have seen a complete change in the quality and quantity of photometric time-series data and we are facing to a revolution of big data in this field. The  satellites MOST \citep{most2003}, CoRoT \citep{corot2006a}, Kepler \citep{kepler2010a} and TESS \citep{tess2014} have been dedicated to photometric time-series observations from space and delivered hundreds of thousands public light curves with cadence as low as 1 minute. We went from \textbf{$\sim8300$ signal-to-noise per 1 minute integration on a $m_V$ Sun-like star ($\sim120$ ppm for MOST satellite) to $\sim40$ ppm for a $m_V=7$ Sun-like star (Kepler) with 1 minute exposure \citep{keplerhandbook}.} TESS is observing the entire sky and plan to deliver 200,000 light curves of bright stars. \textbf{The PLAnetary Transits and Oscillations of stars} (PLATO) \citep{rauer2014} will deliver light curves with 25 seconds cadence of an extremely wide field ($\sim$2232 deg$^2$). Accurate use and manipulate this huge quantity of light curves and measure observables \textbf{could be} difficult. The majority of the pipelines used for these missions are based on transit signal cut by modelization of the light curve via Markov chain Monte Carlo (MCMC) \citep{dey2015, pr2016} and use all the possible combinations of eccentricity, inclination, luminosity ratio and other physical parameters that influence the overall aspect of the light curve. In many cases, as the modelization uses all the light curve signal, the spot modulation sometimes, is suppressed within the removal of the transits from the light curves. Other possibility to subtract the eclipsing binary transit is to detect lower level signals from planets. This refines the orbital period of the binary to better erase the trend of the light curve \citep{corot2017}, and in \textbf{this} process, \textbf{rotational modulation signal can also be erased.} The characterization of the physical parameters presented in most codes to analyze binary system use an approach that does not enable the separation between spot modulation signal and the signal from the binary transiting system. \\

The differential method we propose here consists to identifying variations greater than a cutoff threshold in the time computed derivative of the entire light curve (LC). Thus, \textbf{from that signal with high variations identified we have} a new time series that represents the net variation of the curve. This method makes it possible to disentangle the signal from the transits from other generic signals, and to make a separate analysis of both stellar rotation and transit signal components. This method will be beneficial, for example, to PLATO  mission  \citep{rauer2014}. This is a space mission from the European Space Agency (ESA) whose science objective is to discover and characterize new extrasolar planets and their host stars. PLATO is Expected to be launched by the end 2026. This mission will focus on finding photometric transit signatures of Earth-like planets orbiting the habitable zone of main-sequence Sun-like stars, however, besides the exoplanetary transits, thousand of eclipsing binaries will be also observed. Thus, this is a preparatory study to measure stellar rotation  for stars with companions and to be applied to light curves  from CoRoT, Kepler, TESS as well to the  PLATO mission (ESA).

\section{Eclipsing binaries light curve} 

\begin{figure*}[t]
\includegraphics[width=0.9\textwidth]{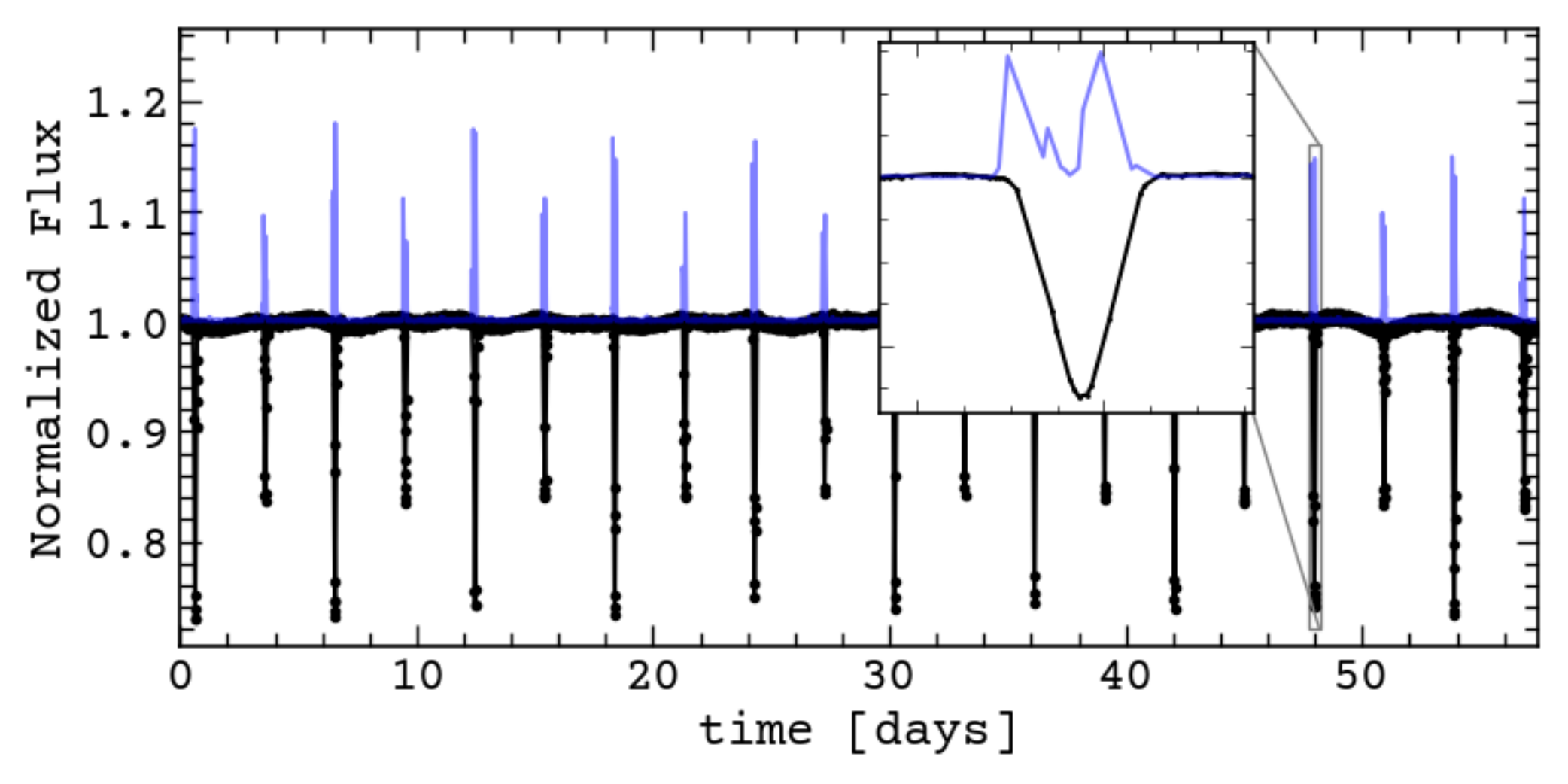}
\caption{Black line represents the raw light curve of CoRoT 102681270 extracted from idoc-corot.ias.u-psud.fr. Shaded blue line is the time series originated from the DRUM TONES routine and representing the differential variation of the raw light curve. The inside plot is a zoom of one region and showing in detail the differential curve overlapping the transit signal.}
\label{method}
\end{figure*}

Eclipsing binary systems are typically characterized through some basic physical parameters that affect the shape of its light curve. The orbital period of the system can be identified from the time span between the primary eclipsing signal. The relative radius of each star can be inferred from the relative depth of each transit signal in the light curve, \textbf{and quantities as eccentricity and inclination can be better constrained when we use both light curve and radial velocity analysis}. Mass and radius are commonly represented in units of solar mass (M$_{\odot}$) and solar radii (R$_{\odot}$). It is expected that the spot modulation, presented on cool stars of \textbf{detached} binary systems, follows the same physics as to our Sun. The regions of increased magnetic activity constrain and cool the encapsulated gas, resulting in a region of lower temperature, and thus given a darker region appearance. These regions cause asymmetries in the light curve and, depending on the latitude, size, and duration, give us an indication of the stellar surface rotation. The typical eclipsing binary we are interested in is the low cadence detached system. \textbf{Such a} system gives a typical light curve with some standard features as shown by black line in Figure \ref{method}.\\

\textbf{Typically this} system reaches the main sequence with their stars spinning faster than the orbital rate. As they are close, the tidal force acts in the components and the torques can play a major role than the rotational braking due to magnetic coupling with the stellar wind, and the orbital angular momentum can dominate and control the stellar rotation as described by \cite{Massarotti2008}. The tidal interaction allows the systems to reach the state of minimum kinetic energy, through the exchange of angular momentum between the stellar rotation of the components of their orbital motion. In this state\textbf{,} the orbit is circular, the rotation of both stars is synchronized with the orbital motion, and their spin axis is perpendicular to the orbital plane, as described by \cite{Zahn2008}. \textbf{In this case,} the standard procedure to characterize a binary system is to complete modelization \textbf{of} the system and then an extraction of the modelized light curve from the original one. In most cases, that procedure ends up erasing weak signals that may contain stellar rotation information from spots \citep{2019MNRAS.489.1644W}.

\section{The DRUM TONES algorithm}

\begin{figure*}[t]
\includegraphics[width=0.9\textwidth]{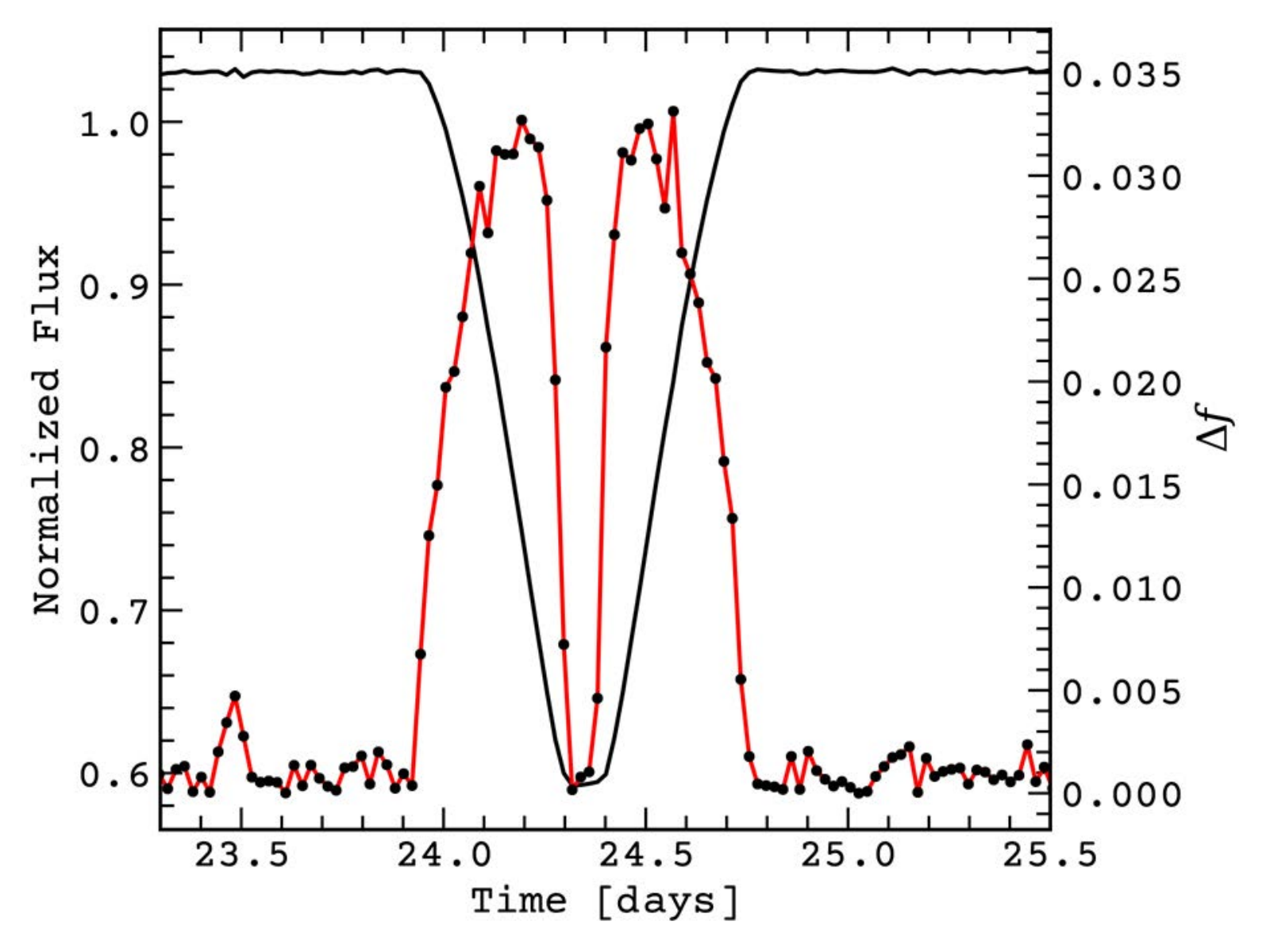}
\caption{The variation time series generated by the DRUM TONES routine representing the differential variation of the raw light curve (solid red line with black points). The solid black line is the eclipse signal of the primary star.}
\label{methodzoom}
\end{figure*}

The DRUM TONES algorithm (DT) was developed to identify as accurately as possible, the transit times\textbf{,} if any exists. The algorithm identifies the beginning and the end of the eclipses in any given light curve. Technically, the DT identifies where happens the greatest variations of flux in a complete light curve. By \textbf{estimating} the differential of the light curve $df/dt$, we obtain a new time series that represents the abrupt variation of the light curve. On the other hand, we also compute the variation of each flux point with respect to the latter point and the next point. The variation spectra can be generated by using the equation $F_{i} = |f_{i+1} - f_{i}|$ and $\Delta F = [F_{1}, F_{2}, F_{3} ,..., F_{n}]$ where $\Delta F$ is the new time-series that contains the differential fluxes, $n$ is the length of the time-series and $f_i$ is the singular flux value in each index $i$. If we are dealing with a \textbf{low cadence EB} light curve, the greatest variations on the time series will indicate the beginning and the \textbf{ending} of each eclipse as shown in Figure \ref{methodzoom}. \textbf{From Figure 1 we can see the spikes in blue matching the eclipses positions.} \textbf{This allow us to identify} the beginning, the middle and the ending of \textbf{each} eclipse. \textbf{The} time-series variation will be characterized as the region that contains the two big spikes, with low variation in the middle. That represents the eclipse regions in time. \textbf{In our analysis, we can also determine the differential} time series to show, in the inflection points, the direction of growth of the function, which gives us mathematically the beginning and \textbf{ending} of the eclipse. As this process identifies all points out of a smooth tendency of the time series, we can also apply a threshold to it and construct a mask to identify the largest variations along with the \textbf{derivative} time series. After that, the threshold mask, \textbf{originally created on the derivative time-series} is then applied to the \textbf{original} light curve. \textbf{This is different from the sigma-clipping method, which consists in clipping all data-points of the light curve that lie more than a fixed $\sigma$, which is, most times, the mean observations error.}\\

By applying the threshold mask to all light curve, we eliminate all transits signals, and we keep discontinuities all over the light curve where all the transit signals should be. These '\textbf{gaps}' generate false positives periods from Lomb-Scargle as showed by \citet{vio2013} and \citet{vanderplas2018}. At this stage is important to fill these \textbf{gaps} with points along the transit.  \textbf{By} fitting the curve gaps, we can bring back a residual time series, that after some adjustments represent the spot modulation of the EB, as shown in the first panel of Figure \ref{DT_good_analysis}. At the end of this process we,\textbf{,} have the model of the spot modulation (Model) without the transit signals and other discrepant points. By subtracting the model from the original light curve, we get a new curve that contains all the original signal\textbf{,} but without the spot modulation. And finally, by analyzing the light curve without the transit signals from binaries and/or planets, we can \textbf{measure} the stellar rotation period \citep{basri2011}. 

\subsection{Basic usage}\label{basicusage}

\begin{figure*}[t]
\includegraphics[width=0.9\textwidth]{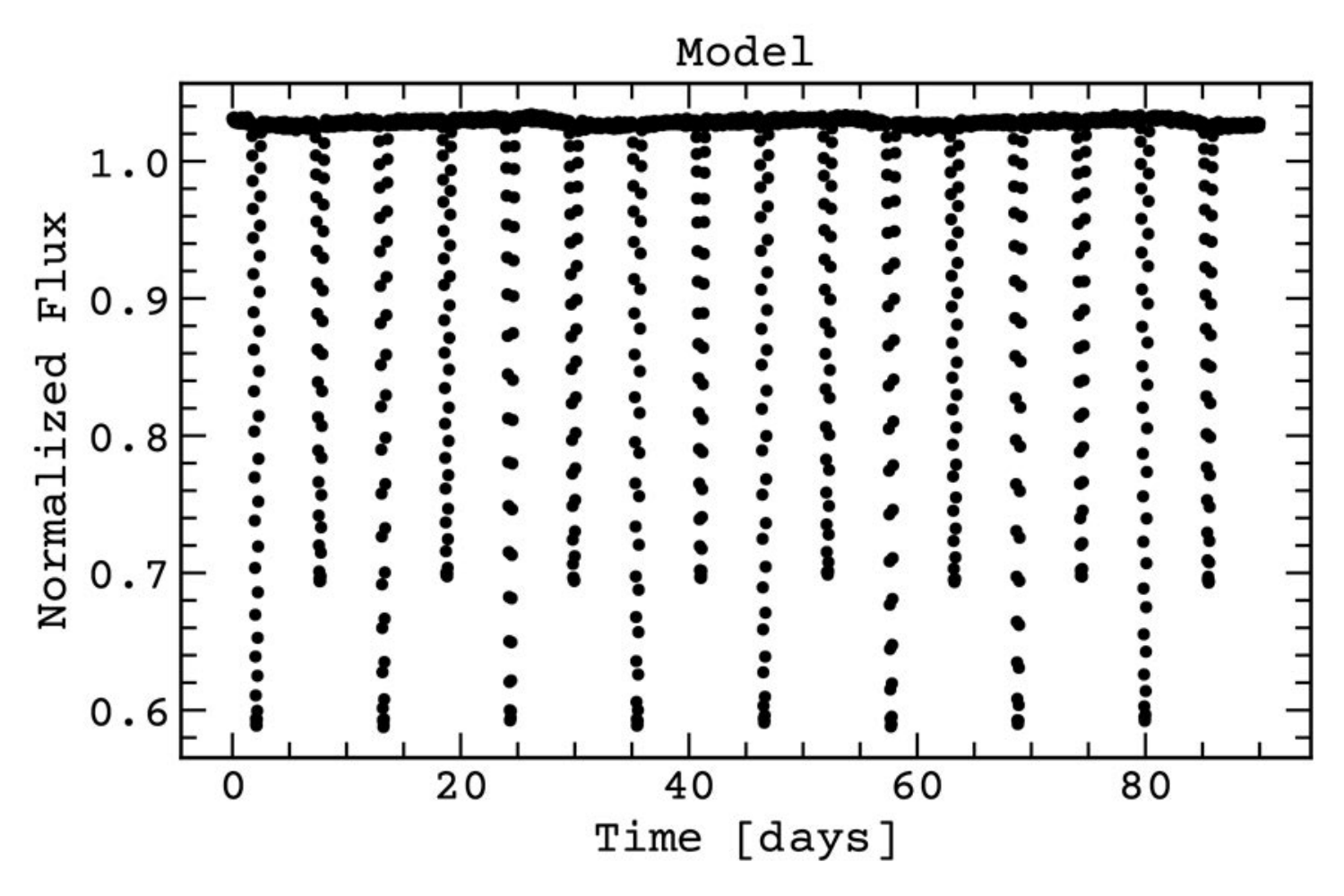}
\caption{Raw modeled light curve with the parameters listed in table \ref{model_good_table}.}
\label{DT_good_raw}
\end{figure*}

In order to apply the DT code in a controlled experiment of a binary system, we used the \textbf{1.6.1} version of the $ellc$ light curve model \citep{Maxted2016} to simulate an eclipsing binary light curve with the primary star of the system presenting spots modulation and limb darkening profile. First, we created \textbf{a} spotted star with parameters presented in table \ref{model_good_table}, then we created the binary system with both stars as spheres and \textbf{both} with the same limb darkening coefficients as a rotating star, also presented in table \ref{model_good_table}. \textbf{The final signal is the composition of both light curves}. We then applied a typical Gaussian error of about $1 \times 10^{-3}$ of the total flux signal. The final simulated light curve is presented in Figure \ref{DT_good_raw}. At first glance, we can see that this light curve presents both eclipsing binary and spot modulation signals, however it is not an easy task to disentangle the rotation modulation from the transit signal. The common step from most procedures, starts by the detrend of the spot modulation signal to get out only the binary transit, however, for some applications, it is important to maintaining rotational modulation of the primary component of the binary system. This is the typical \textbf{example} were our algorithm could be applied. \\

\begin{table}[]
\caption{Physical parameters for $Model$ synthetic binary system. $ldc_1$ and $ldc_2$ are the limb darkening coefficients for a squared law. $r1$ and $r2$ are the fraction radius of the primary and secondary star of the synthetic binary system.}
\begin{tabular}{cccccccccc}
\begin{tabular}[c]{@{}c@{}}$P_{rot}$\\ {[}days{]}\end{tabular} & \begin{tabular}[c]{@{}c@{}}Spots\\ {[}\#{]}\end{tabular} & ldc\_1 & ldc\_2 & SB\_ratio & \begin{tabular}[c]{@{}c@{}}Incl \\ {[}degrees{]}\end{tabular} & r1   & r2   & \begin{tabular}[c]{@{}c@{}}$P_{orb}$\\ {[}days{]}\end{tabular} & \begin{tabular}[c]{@{}c@{}}Cadence\\ {[}min{]}\end{tabular} \\ \hline
        &      &       &       &       &       &       &       &        &       \\
27.5    &3     & 0.5   & 0.5   & 0.45  & 89.5  & 0.13  & 0.10  & 11.12  & 30    \\
        &      &       &       &       &       &       &       &        &       \\ \hline
\label{model_good_table}
\end{tabular}
\end{table}

\begin{table}[]
\caption{DT basic input and description}
\label{inputs}
\begin{tabular}{cl}
Input       & Description                                                                                                                                                                                                                 \\ \hline
ID          & is the target identification that must be in the same folder as the code                                                                                                                                                    \\
time        & is the array that contains the time of the light curve                                                                                                                                                                     \\
flux        & is the array that contains the flux of the light curve                                                                                                                                                                      \\
err         & is the array that contains the flux error                                                                                                                                                                                   \\
detrendy    & is the polynomial degree to be fitted to the residual spot modulation                                                                                                                                                       \\
splitfactor & is the number of sectors to split the light curve to be processed individually                                                                                                                                              \\
condition1  & \begin{tabular}[c]{@{}l@{}}is the max variation possible at the variation space to be accounted as out of \\ the threshold\end{tabular}                                                                                     \\
cadence     & \begin{tabular}[c]{@{}l@{}}is the change on the cadence, if this is necessary, 1 means no change, 2 means \\ that the light curve will have now be binned with half of the number of points \\ than before\end{tabular}     \\
condition2  & \begin{tabular}[c]{@{}l@{}}is the limit on the time-space between discontinuities, (e.g the minimum time to \\ consider a change of quarter of about 90 days) normally is greater than the \\ original cadence\end{tabular} \\ \hline
\end{tabular}
\end{table}

The DT algorithm is not a fully automatic code and needs to be fed with some custom parameters. In order to use it, we need to specify inputs as described in table \ref{inputs}. The DT algorithm is available on the Zenodo Repository\footnote{(DRUM TONES, Version 1.0.0). Zenodo. http://doi.org/10.5281/zenodo.1472861} \citep{dealmeida2018} and it is ready \textbf{to go} if the user has installed the Python 3.x and the libraries matplotlib and numpy.

\begin{figure*}[t]
\includegraphics[width=0.47\textwidth]{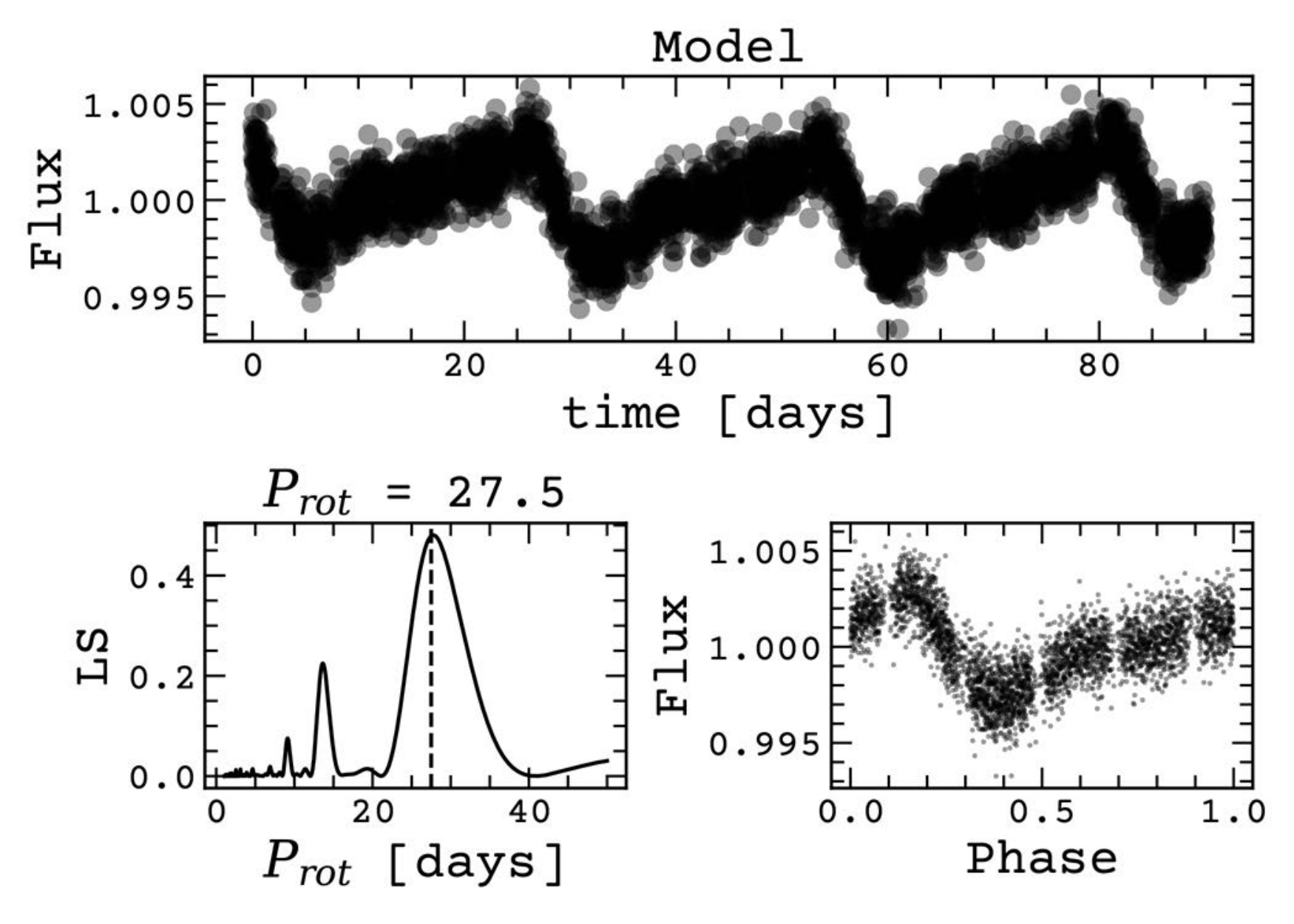}
\includegraphics[width=0.47\textwidth]{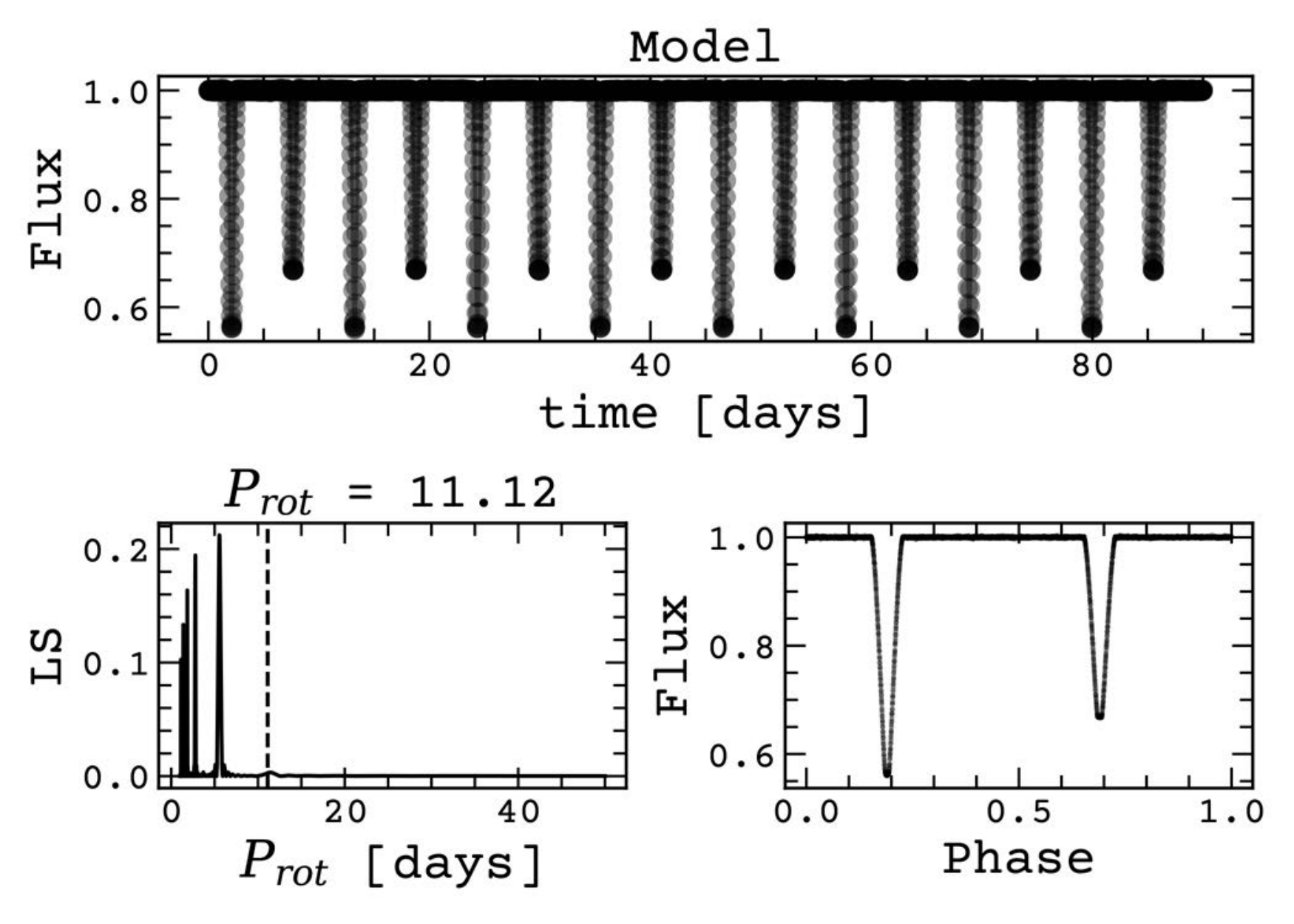}
\caption{Simulated eclipsing binary light curve with spot modulation and Gaussian error. Left: Rotational modulation on top. Lomb-Scargle periodogram, and phase \textbf{diagram} at the \textbf{measured} period at the bottom. Right: Eclipsing binary system analysis following the same panels as the rotational modulation analysis.}
\label{DT_good_analysis}
\end{figure*}

\section{Space Missions applications}

Even if our goal was to develop an algorithm to be used along with the PLATO mission, we tested our procedure by analyzing published light curves from the space missions CoRoT, Kepler, Kepler K2, and TESS. These missions produced a rich and large time-series survey with years of time coverage high-quality data. We also generated and used a sample of synthetic light curves for the coming space mission PLATO\textbf{,} simulated with intrinsic noise as described by \citet{plato2019a}. Details about the amplitude of instrumental systematics and accuracy of some used parameters are described in the following sections. \textbf{The} results for the analyzed light curves are presented in table \ref{alltable}.

\subsection{CoRoT Mission}

\begin{figure}[h]
\includegraphics[width=0.47\textwidth]{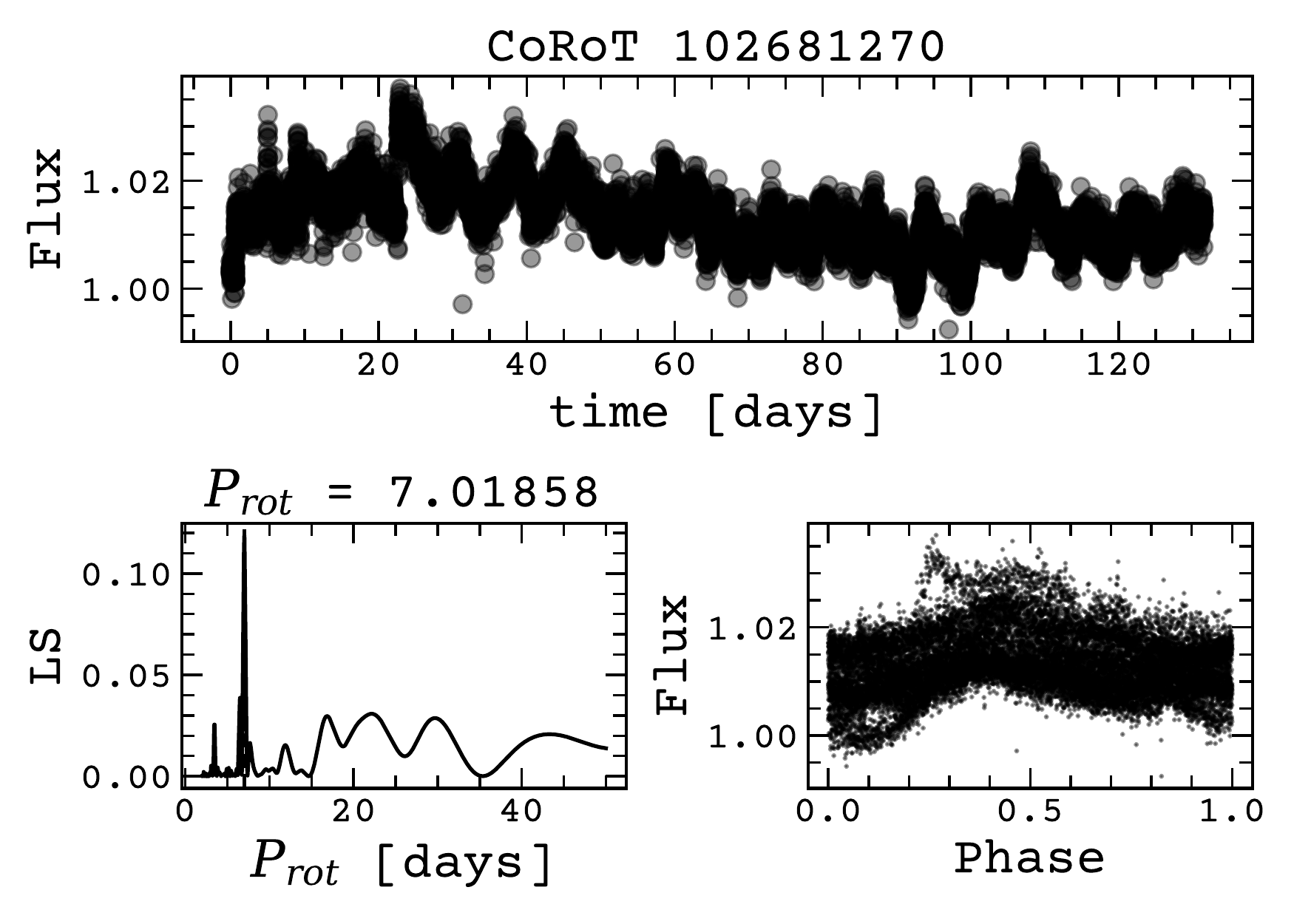}
\includegraphics[width=0.47\textwidth]{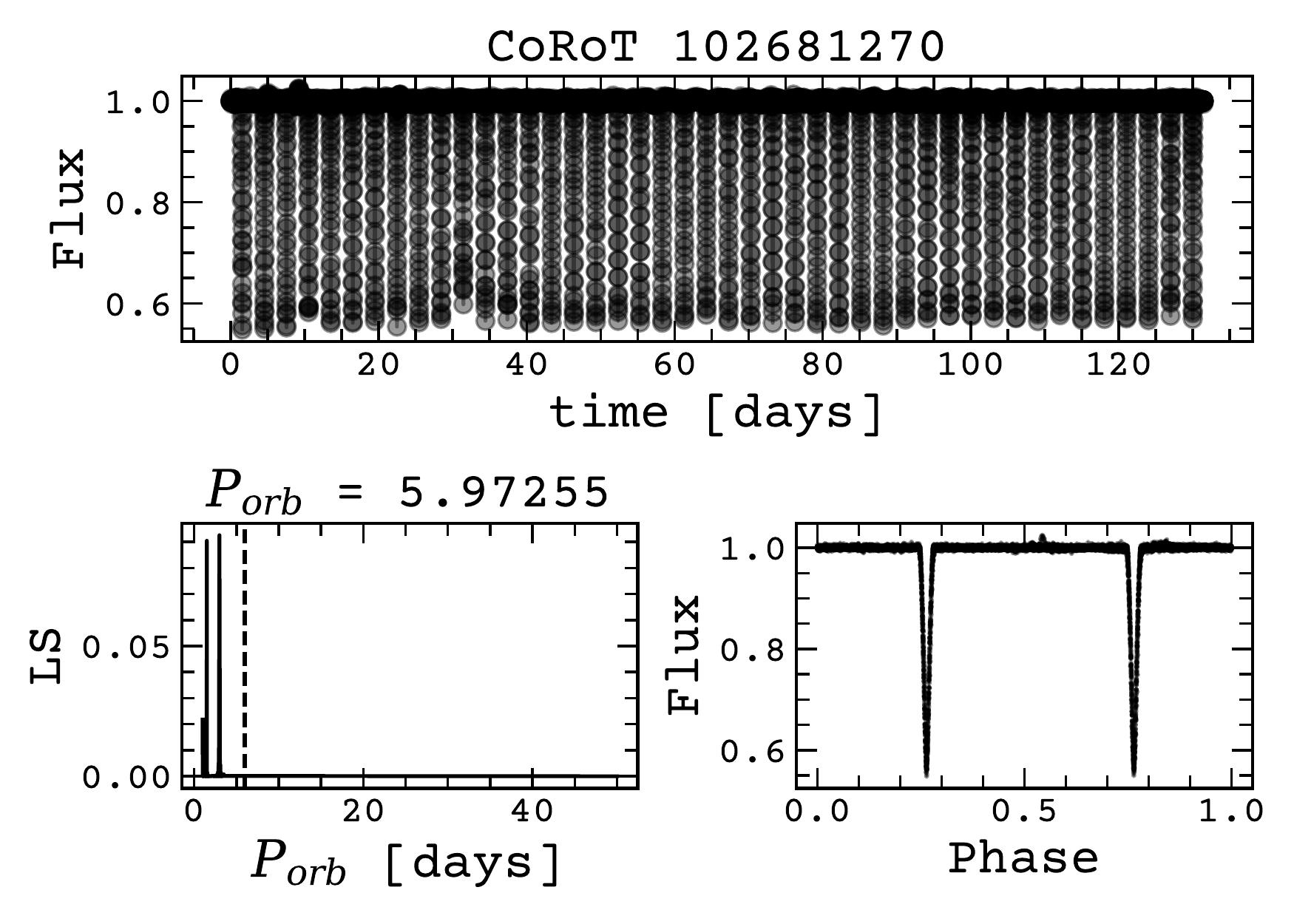}
\caption{CoRoT 102681270 light curve without trend correction. Left: Rotational modulation on top, Lomb-Scargle periodogram, and phase at the found period at the bottom. Right: Eclipsing binary system analysis following the same panels as the rotational modulation analysis.}
\label{COROT_analysis}
\end{figure}

The CoRoT (CNES) mission, launched into a low-Earth polar orbit on 2006 December 27, \textbf{monitored} several thousand stars at once with combined four 2048 x 4096 pixel EEV CCDs. The field of view is a square of 2.8 x 2.8 degrees. CoRoT \citep{2009A&A...506..411A} was, therefore, the first mission focused on searching for exoplanet transits from space. The CoRoT telescope has a diameter of 27 cm and stored the light curves in 3 bands: red (R), green (G) and blue (B) obtained through the insertion of a low-resolution prism in the light path focus of the telescope. However, these bands do not correspond to the photometric filters and are different for each star \citep{corot2006}. When dealing with the CoRoT light curves, we need to specify which color to use or apply the total integration of all colors, defined as white band (W). For each color file or for the white integrated file, a flag corresponding to the quality of the measurement is presented. As the 3 bands are different, we end up with a light curve with a bunch of problems, as for example discontinuities and jumps. CoRoT observed two regions of the sky, one towards the Galactic anticenter ($RA2000=$06h50m25s, $DEC2000=$-01 42 00) and the other towards the Galactic center (RA2000= 19h23m34s, $DEC2000=$ 00 27 36 ) \citep{corot2012}.\\

An eclipsing binary system observed by CoRoT is the \textbf{star} CoRoT 102681270 (2MASS J06431548-0040407). This system has orbital period of 2.987 days, as published \citet{corot2012}. However, probably due to the difficulties described here, these authors have not published the rotation period of the primary component. This object presents a light curve with a strong transit signal and some systematic issues. We applied the DT algorithm in this object, and we were able to disentangle the apparent spot modulation from the transit signature. The top first panel of Figure \ref{COROT_analysis} shows the spot modulation extracted from the original light curve. We can see that, \textbf{despite this light curve presenting some systematics}, the algorithm correctly identified the rotation and the orbital period transit signal. For CoRoT 102681270, the measured rotation period was \textbf{7.02 days}. \textbf{As the primary and secondary transits have almost the same depth, aliases in the Lomb-Scargle periodogram get masked and the measure of the orbital period by the peak in the Lomb-Scargle actually reflects half of the real orbital period}. CoRoT 102681270 example shows that most likely, despite having done a great job, the authors above mentioned have measured half orbital period aliases for this object and do not reach measurable rotational period value. In Figure \ref{COROT_analysis} we summarize all measurements for CoRoT 102681270.

\subsection{Kepler and K2 Mission}

\begin{figure}[h]
\includegraphics[width=0.47\textwidth]{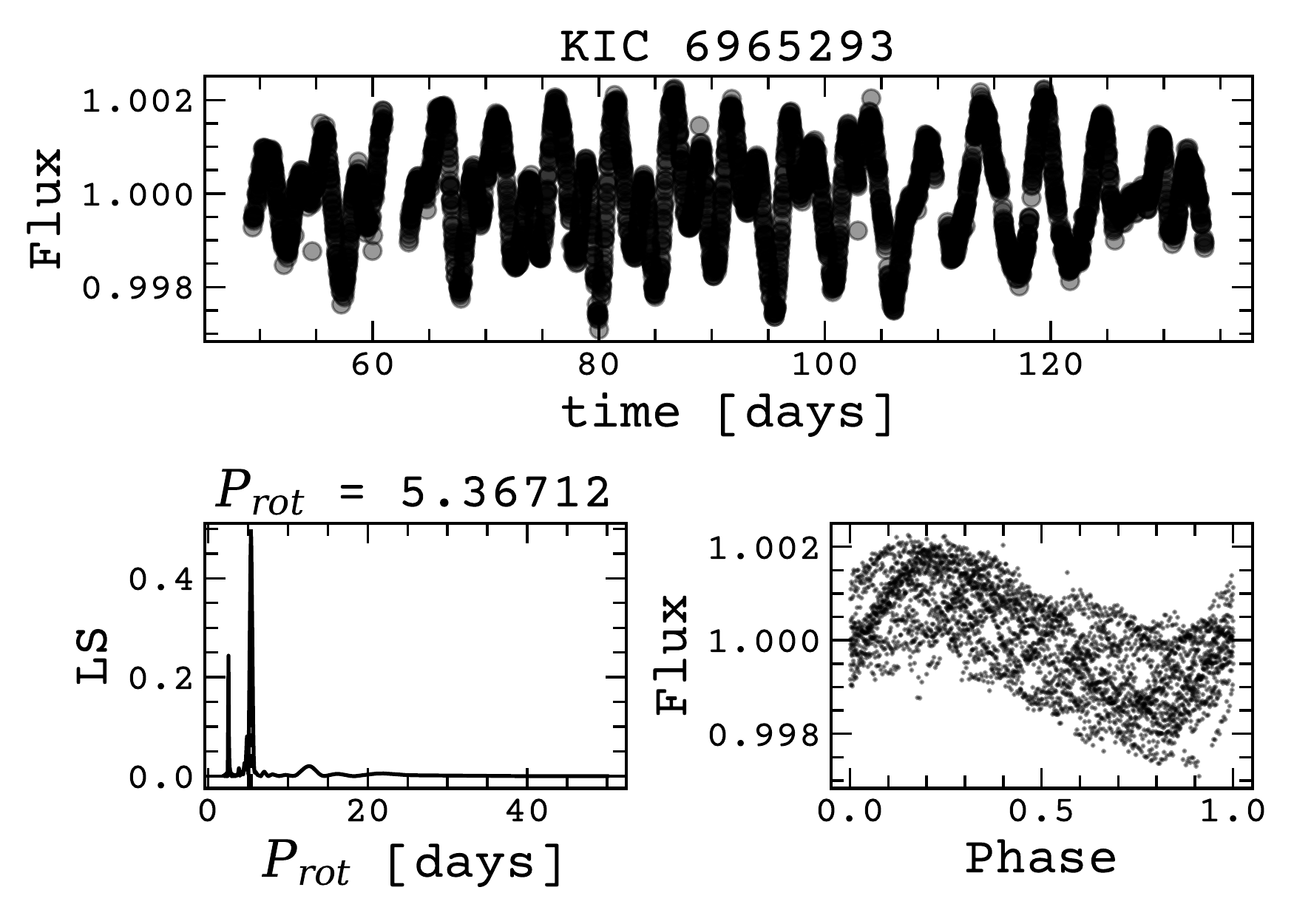}
\includegraphics[width=0.47\textwidth]{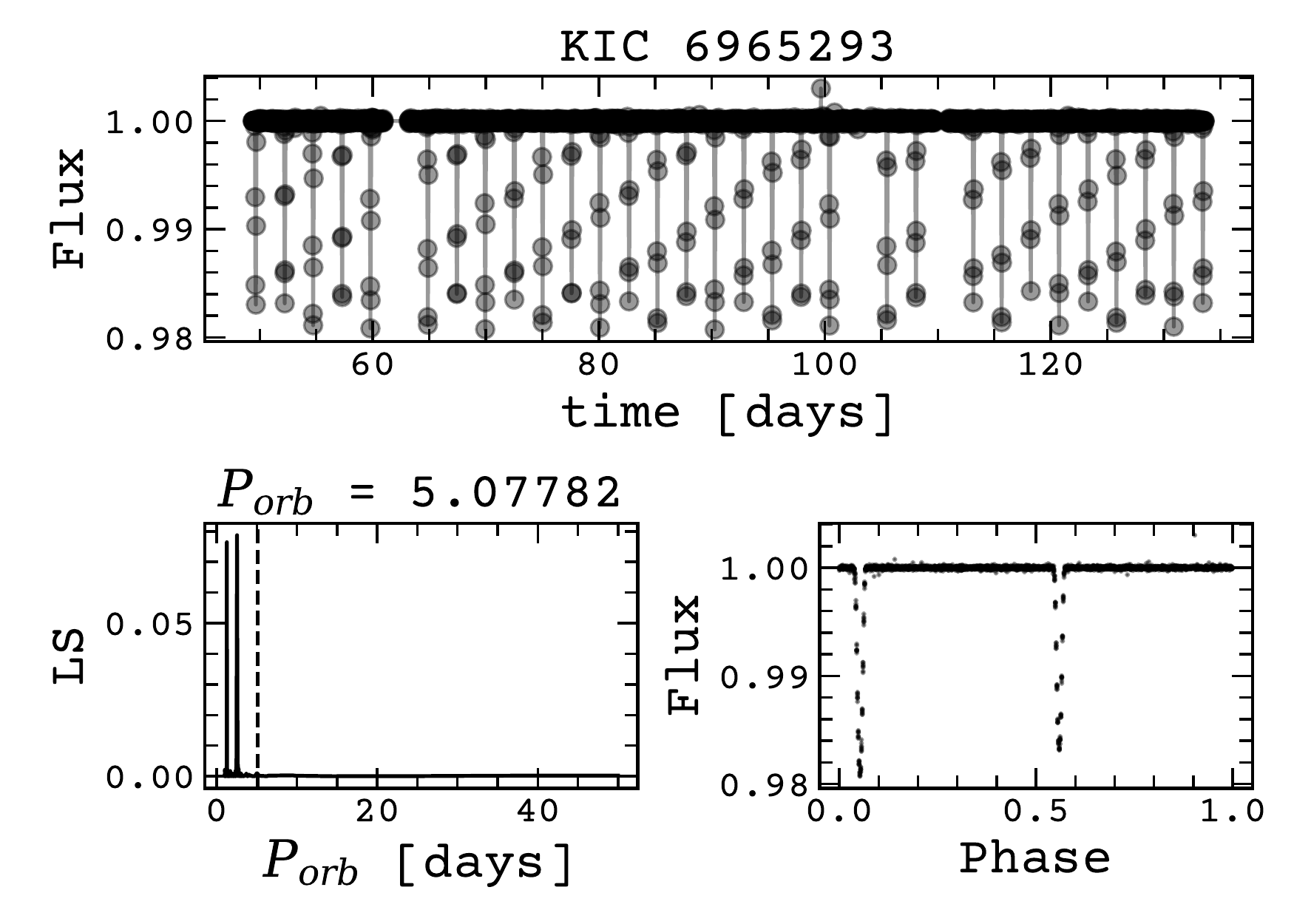}
\caption{KIC 6965293 light curve from the second quarter of the Kepler mission. Left: Rotational modulation on top, Lomb-Scargle periodogram, and phase at the found period at the bottom. Right: Eclipsing binary system analysis following the same panels as the rotational modulation analysis.}
\label{KEPLER_analysis}
\end{figure}

Another application we focused was based on eclipsing binary candidate from Kepler mission. During $\sim$4 years of operations, the Kepler space mission delivered continuous, high-precision light curves for over 150,000 stars, with a cadence of 29.4 min. The Kepler light curve files contain two versions of the light curve. They are SAP (simple aperture photometry) and PDC (pre-search data conditioning). \citet{2010SPIE.7740E..0DJ} and \citet{fanelli2011kepler} give an overview of the data processing steps involved in producing both the target pixel files and the light curves. The PDC light curves result from more advanced processing steps designed to remove instrumental artifacts and systematics \citep{2012AAS...22033003S, 2012PASP..124..985S}. Thus, the PDC pipeline is primarily ready data for planetary transit searches, as well \textbf{as} to study astrophysical photometric variability from the observed stars. The final version of the Kepler data we used was available directly from the MAST web site and uses the PDC-msMAP pipeline. For a description of systematics, impact and removal process on the Kepler light curves, see e.g., \citet{2017MNRAS.471..759A}. \textbf{Based on Kepler data \citet{pr2011} defined these EB systems into five} five groups: detached, semi-detached, over-contact, ellipsoidal and uncertain. The catalog provides information about the orbital period, time of the eclipse, classification, galactic longitude and latitude, Kepler magnitude, effective temperature, long and short cadence data. The major problem on the Kepler data was the discontinuity from corrected light curves with the PDC automatic pipeline.\\

\begin{figure}[h]
\includegraphics[width=0.45\textwidth]{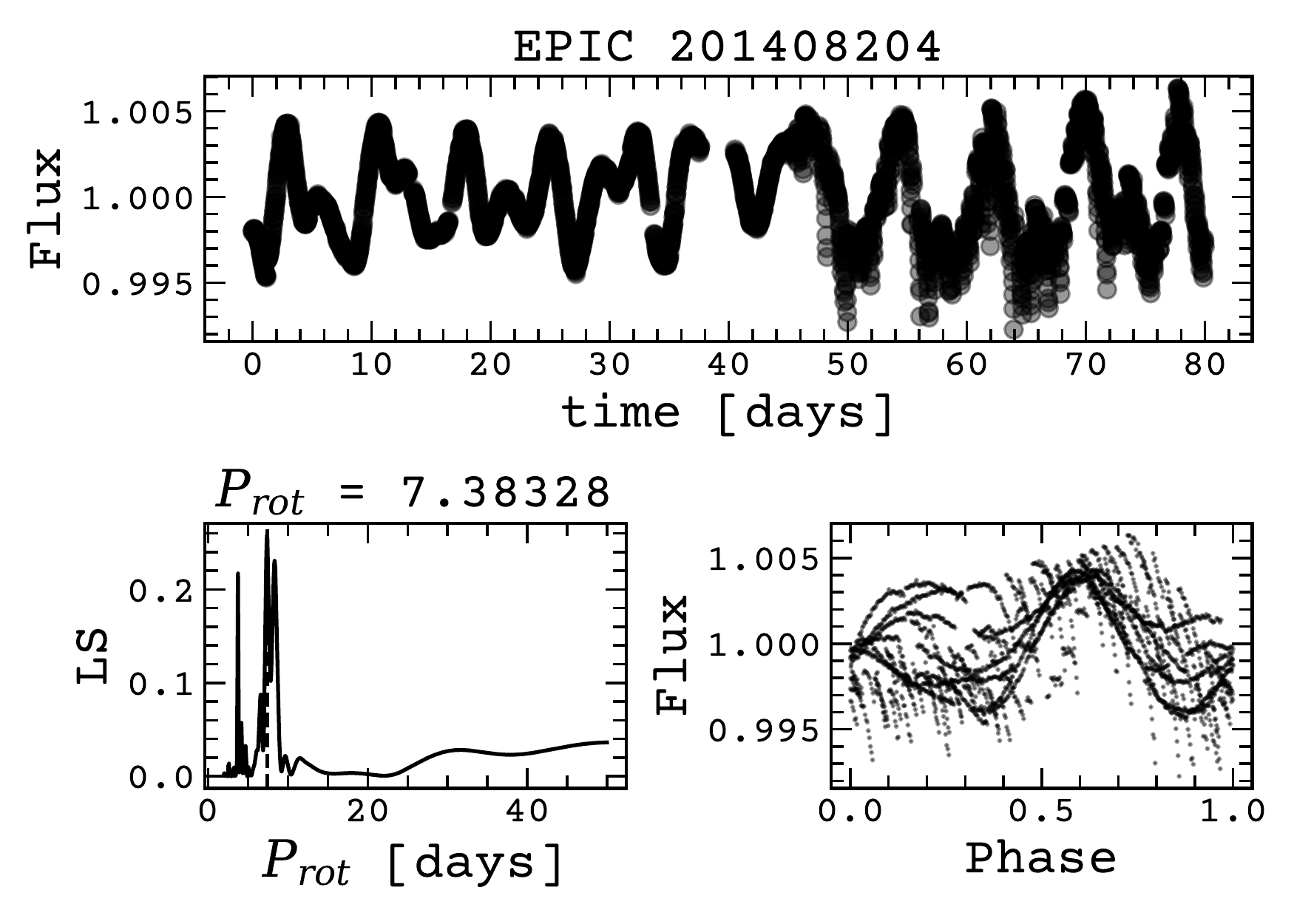}
\includegraphics[width=0.45\textwidth]{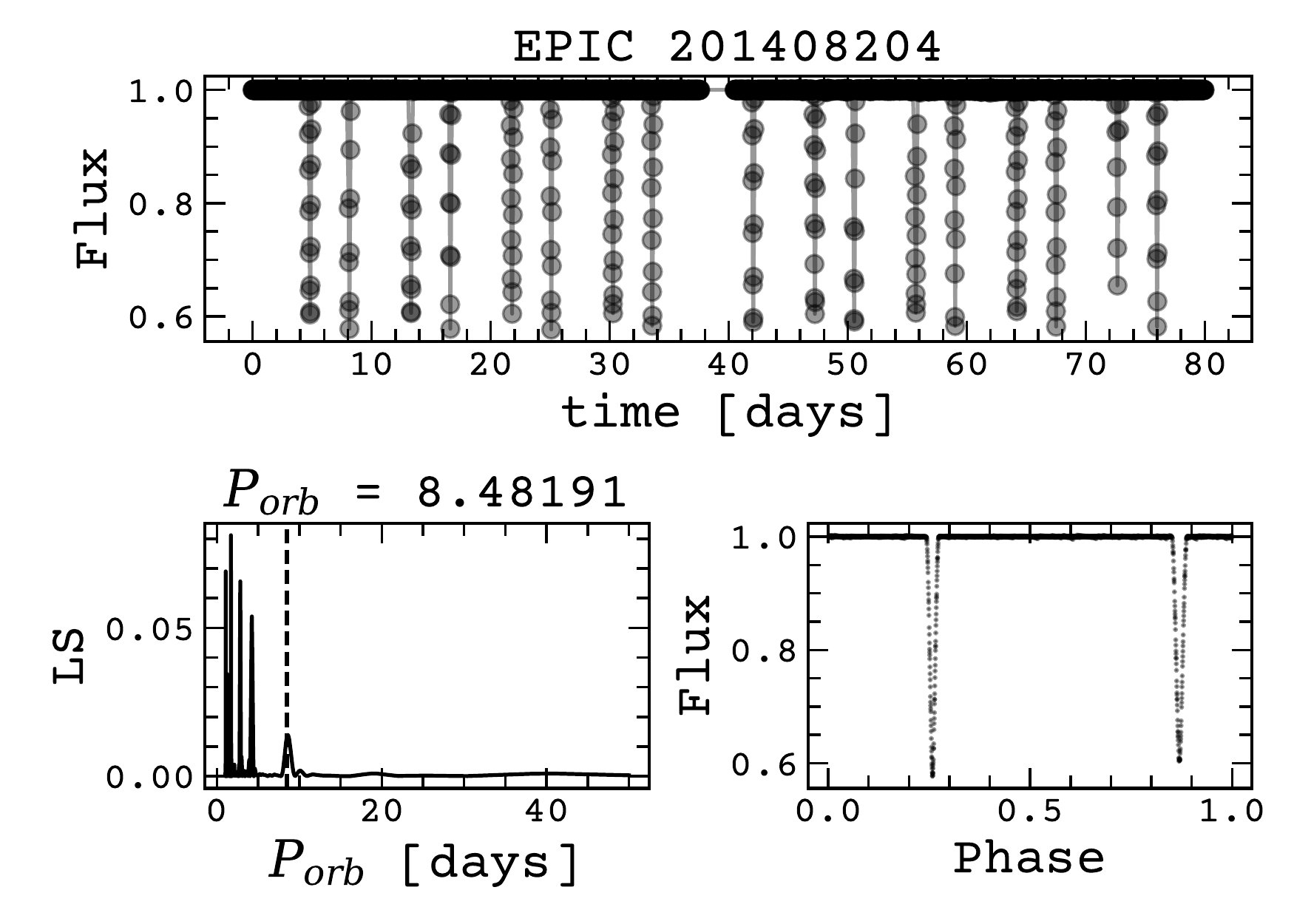}
\caption{EPIC 201408204 light curve from campaign 01 of Kepler K2 mission. Left: Rotational modulation on top, Lomb-Scargle periodogram, and phase at the found period at the bottom. Right: Eclipsing binary system analysis following the same panels as the rotational modulation analysis.}
\label{K2_analysis}
\end{figure}

These quarters flux correction add temporal jumps from scheduled observation. Each of these quarters must be considered separately and then appended to create the final composition of the light curve. This is the standard procedure used by Kepler pipeline. Besides that, for statistical reason and complexity, some minor bad features were still left in the public version of the Kepler light curve data set. These unwanted features have to be removed before any analysis, as well as on eclipse signal treatment. Kepler yielded a catalog of eclipsing binary containing 2878 systems mostly identified by the Transit Planet Search (TPS) algorithm \citep{pr2011}. From Kepler catalogue, KIC 6965293 (KOI-6800) is an eclipsing binary of Algol type with orbital period and classification published by \citet{kepler2011}. This object was observed for 18 quarters and presents a strong spot modulation amplitude. We use the long cadence light curve in the second quarter of the PDCsap data product. The top first panel of Figure \ref{KEPLER_analysis} shows the spot modulation extracted from the original light curve, the second panel shows the binary without the rotation. Panels also show the Lomb-Scargle periodogram and phase diagram results for the rotation and orbital period respectively. The Kepler K2 mission is the extended Kepler mission after one of its wheels fails completely.\\

Operating the spacecraft in this way made it possible to extend the lifetime of the instrument, however, it produces challenging to reduce light curves due to the spacecraft pointing instability. Nevertheless, Kepler K2 is providing high-quality photometry for bright stars in selected regions of the sky named K2 campaign fields near the ecliptic plane \citep{k22014}. Because of the pointing instability of the spacecraft, the resulted light curves are hard to be reduced. This is a perfect experimental situation to use our procedure. There is a variety of algorithm to correct the instrumental noise and recovery the light curve as close to the performance of the original Kepler mission \citep{k22016a, k22014b}. However, most of them also subtract the rotation signal.\\

EPIC 201408204 (TYC 4930-128-1) is an eclipsing binary with orbital and rotational period published by \citet{k22018}. This is an excellent target to test our algorithm because it has both orbital and rotational periods already determined in \textbf{the} literature. We use the systematic corrected light curve modulation from K2 campaign 1, however the overall light curve still presents some systematic problems. The top first panel of Figure \ref{K2_analysis} shows the spot modulation extracted from the original light curve, the second panel shows the binary without the rotation modulation. Both panels at the bottom show the Lomb-Scargle and phase results for the rotation and orbital period respectively.

\subsection{TESS Mission}

\begin{figure}[h]
\includegraphics[width=0.45\textwidth]{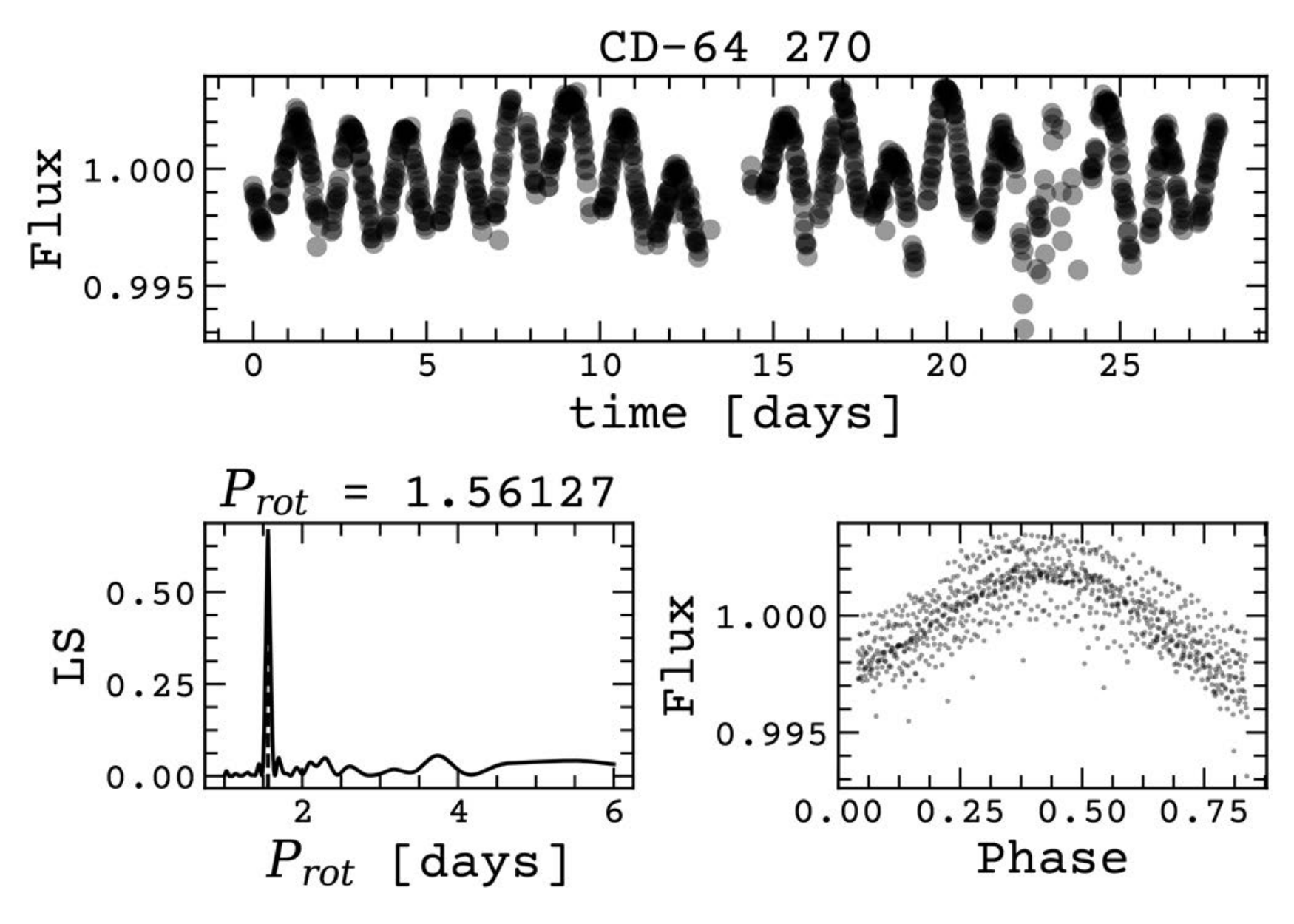}
\includegraphics[width=0.45\textwidth]{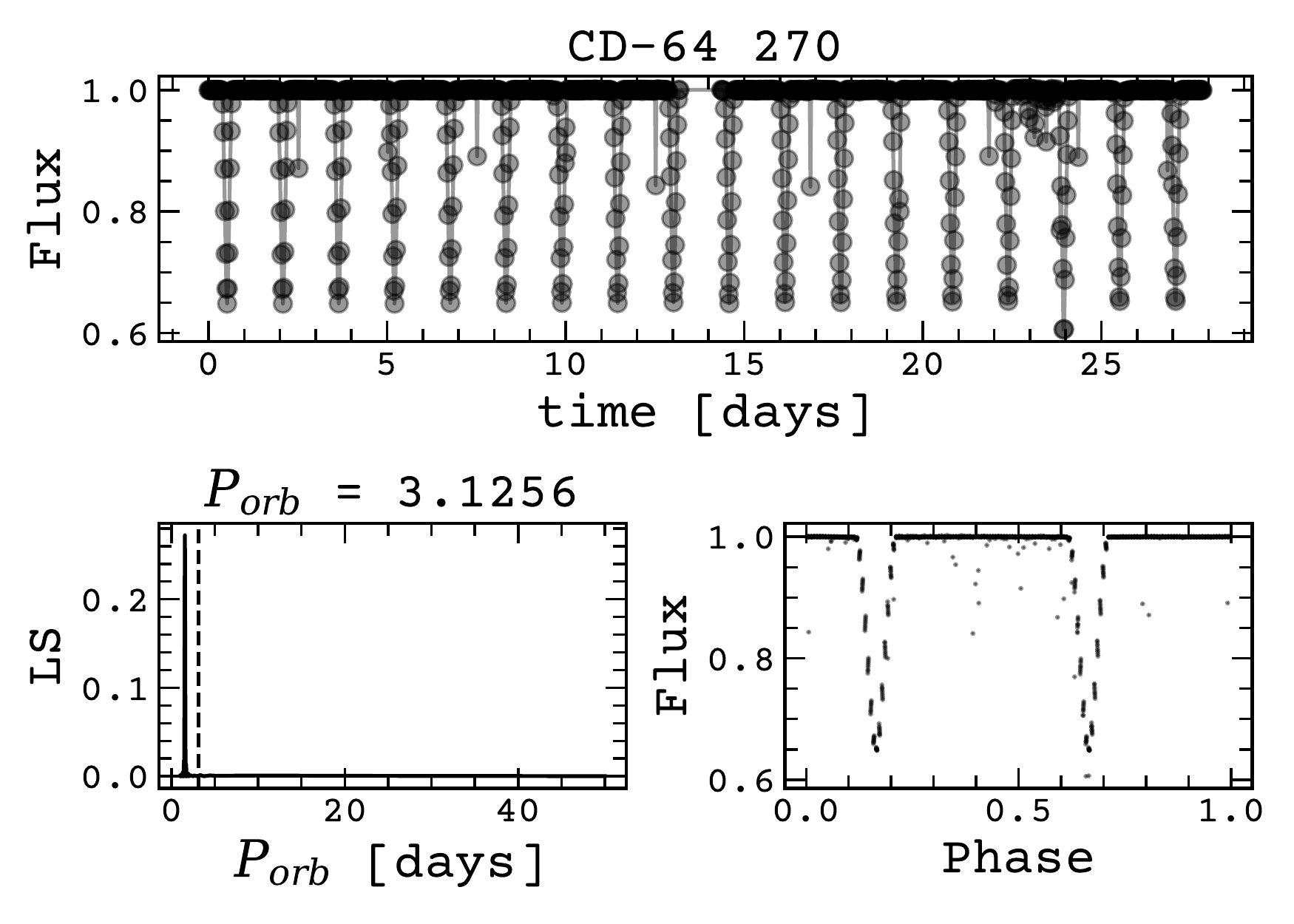}
\caption{TIC 349480507 (CD-64 270) light curve from cam 4 of the 4th sector of TESS mission. Left: Rotational modulation on top, Lomb-Scargle periodogram, and phase at the found period at the bottom. Right: Eclipsing binary system analysis following the same panels as the rotational modulation analysis.}
\label{TESS_analysis}
\end{figure}

The Transiting Exoplanet Survey Satellite (TESS) mission \citep{tess2014}, launched on April 18, 2018, aboard a SpaceX Falcon 9 rocket,  is a 4 camera space telescope with the goal of observing $90\%$ of the sky along two years, and generate light curves for $\sim$150 million stars \textbf{with photometric sensitivity of $\sim$200 ppm (0.02$\%$) for an $I_C$ = 10 star (Cousins I-band centered at 786.5 nm)}. This is enough photometric precision to detect the transits of exoplanets \citep{huang2018,vander2019,wang2019} and enabling stellar astrophysics investigations. TESS light curves will be \textbf{provided} for $\sim$ 400,000 targets observed at 2 minute cadence. However, for most stars will only be provided as full-frame images (FFI) at 30 minute cadence. The TESS image scale of $\sim$ 21 ''/pix is highly susceptible to crowding, blending, and source confusion. Concerning TESS photometric accuracy, \citet{2018AJ....156..132O} show that the pipeline performance does not depend on positions across the field, and only $\sim$ 2$\%$ of stars appear to exhibit residual systematics at the level of $\sim$ 5ppm. Among many objects observed by TESS CD-64 270 (TIC 349480507) is a published eclipsing binary system \citep{tess2019} classified as high proper-motion star in SIMBAD astronomical database. It is a relatively bright star with V magnitude of 9.96. As this object is not a primary target in the TESS mission, we extracted from the Full Frame Images (FFIs) and we correct the systematic in the light curve. The top first panel of Figure \ref{TESS_analysis} shows the spot modulation extracted from the original light curve. The second panel shows the binary with the rotation signal removed. The bottom panes show the Lomb-Scargle and phase results for the rotation period of 1.56 day and orbital period of 3.12 days respectively.

\subsection{PLATO Mission}
\label{plato_subsection}

The accumulated experimental experience with the missions CoRoT, Kepler, Kepler K2, and TESS, gives the motivation to planning a more advanced procedure to be applied for PLATO observations. The PLATO space mission is planned to be launched by 2026/2027 and its goal is to detect and characterize Earth-like exoplanets in the habitable zone \citep{rauer2014}. Even though the main mission of PLATO is to discover exoplanets, it is also expected to detect 10,000 bright EBs with unprecedented quality. At this preparatory time, until the launch of the mission, we did our implementation based on "The PLATO Solar-like Light-curve Simulator" \citep{plato2019a} to generate the systematics of the spacecraft and generate EBs light curves as close as possible to the expected by PLATO. The PLATO mission is far different from  CoRoT, Kepler, and TESS in the matter of high-precision photometry. The mission is based on a multi-telescope concept. The 26  cameras compose the instrument, two of them are named “fast” cameras and will work at 2.5 seconds of cadence while the remaining 24 are named “normal” cameras and work at 25 seconds of cadence.\\

\begin{figure*}[t]
\includegraphics[width=0.9\textwidth]{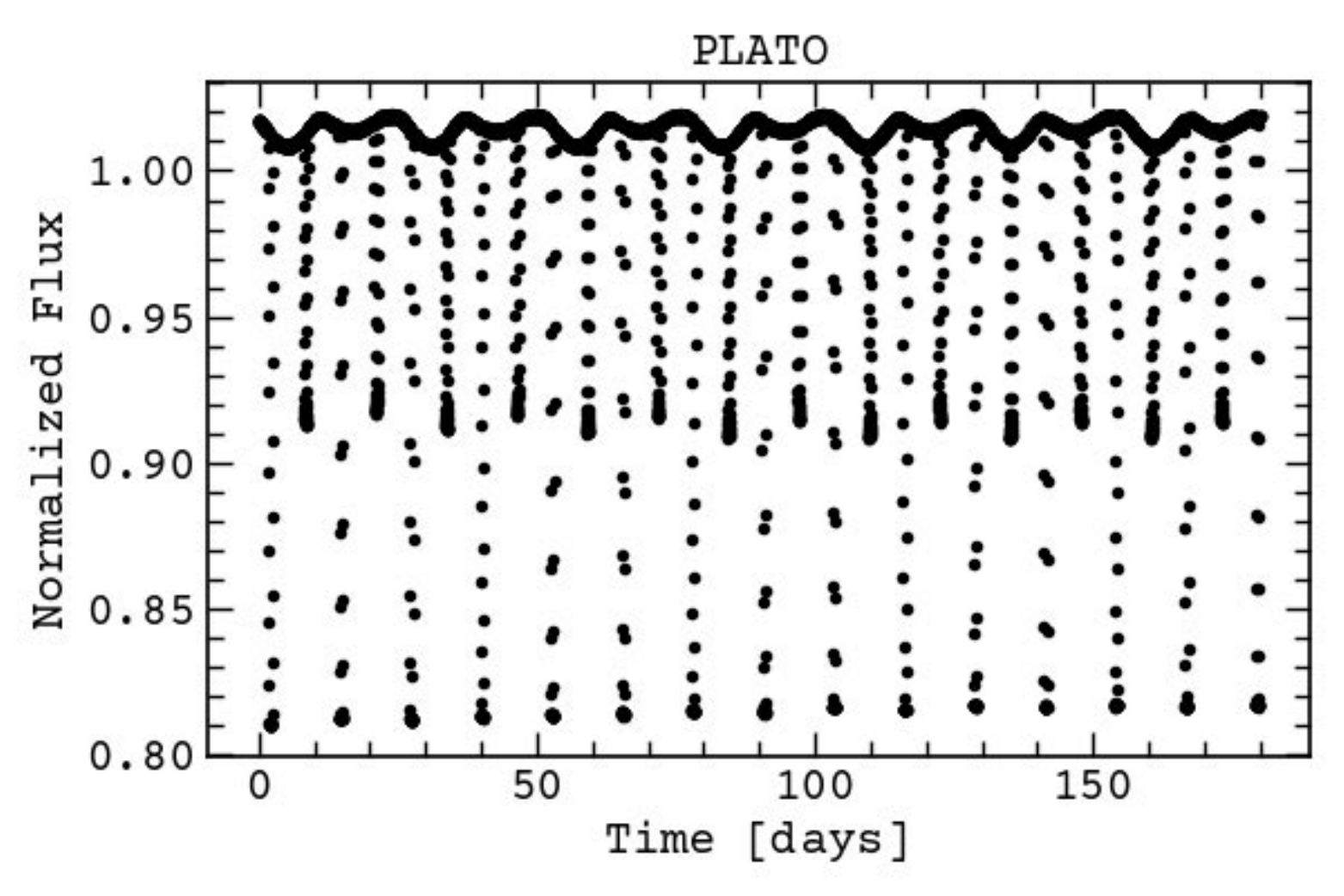}
\caption{Raw modeled light curve with the parameters listed in table \ref{plato_table}.}
\label{rawPLATO}
\end{figure*}

\begin{figure}[h]
\includegraphics[width=0.45\textwidth]{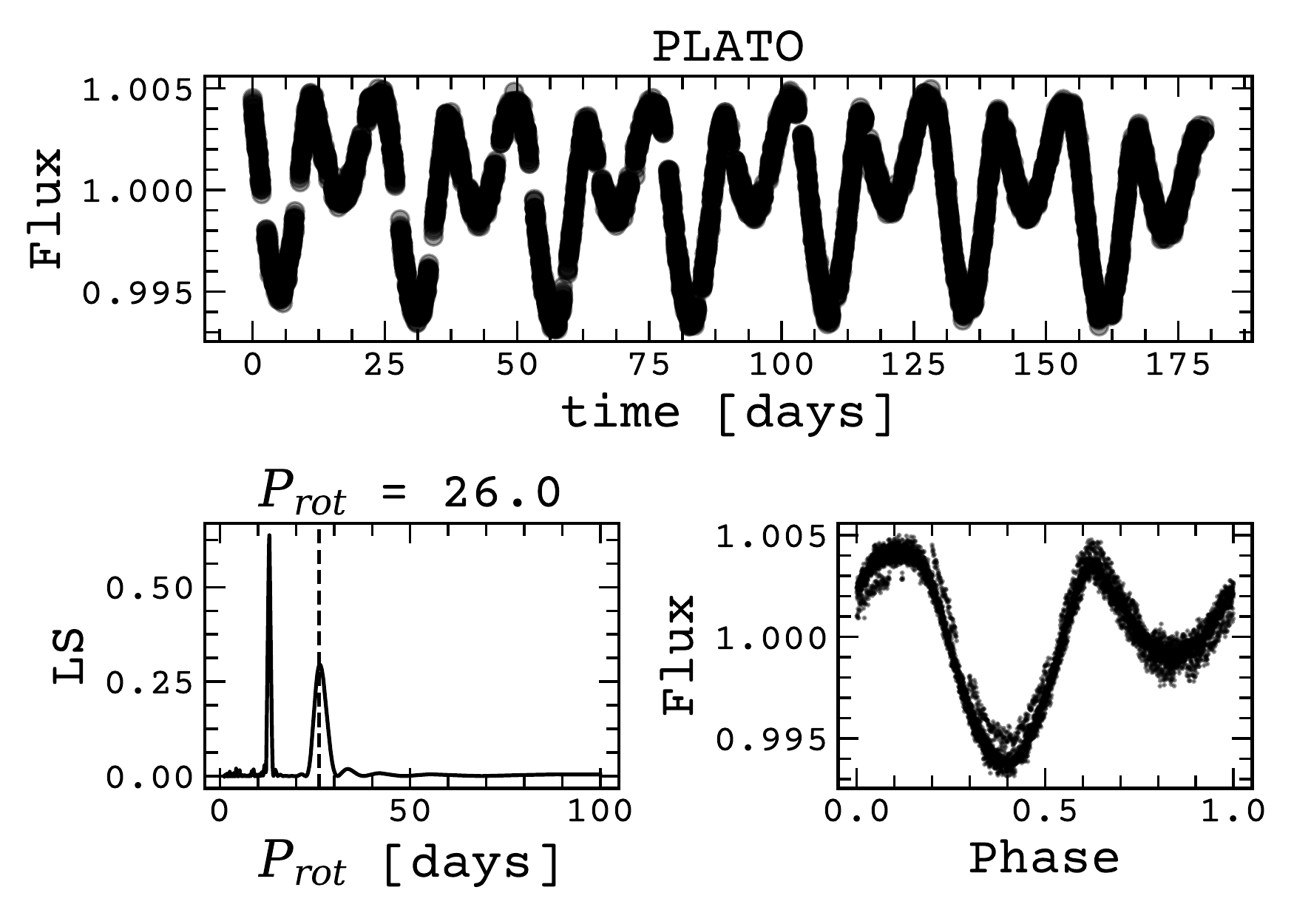}
\includegraphics[width=0.45\textwidth]{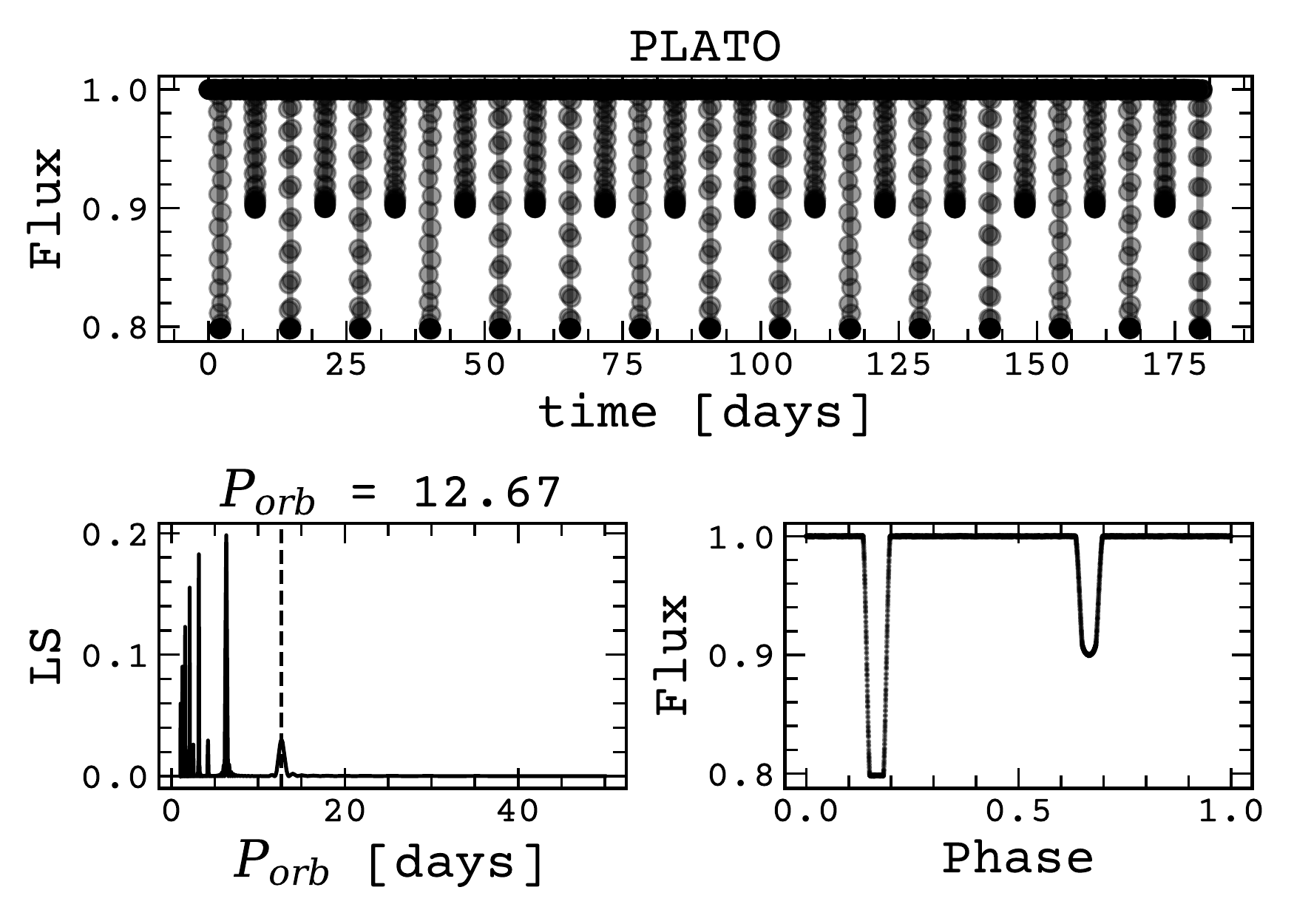}
\caption{Synthetic EB light curve generated with PLATO systematic and rotational modulation. Left: Rotational modulation on top, Lomb-Scargle periodogram, and phase at the found period at the bottom. Right: Eclipsing binary system analysis following the same panels as the rotational modulation analysis.}
\label{PLATO_analysis}
\end{figure}

To simulate our binary system with PLATO systematics, we use the same procedure as described in subsection \ref{basicusage}, and described by a Gaussian error. The parameters of the synthetic light curve are presented in table \ref{plato_table}. The result light curve is presented in Figure \ref{rawPLATO}. After running the DT algorithm in this simulated PLATO light curve, we were able to retrieve the correct rotational and orbital period of the model and orbital period as shown in Figure \ref{PLATO_analysis}.\\

\begin{table}[]
\caption{Physical parameters for $PLATO$ synthetic binary system}
\begin{tabular}{cccccccccc}
\begin{tabular}[c]{@{}c@{}}$P_{rot}$\\ {[}days{]}\end{tabular} & \begin{tabular}[c]{@{}c@{}}Spots\\ {[}\#{]}\end{tabular} & ldc\_1 & ldc\_2 & SB\_ratio & \begin{tabular}[c]{@{}c@{}}Incl \\ {[}degrees{]}\end{tabular} & r1   & r2   & \begin{tabular}[c]{@{}c@{}}$P_{orb}$\\ {[}days{]}\end{tabular} & \begin{tabular}[c]{@{}c@{}}Cadence\\ {[}min{]}\end{tabular} \\ \hline
        &      &       &       &       &       &       &       &        &       \\
26.0    &2     & 0.5   & 0.5   & 0.45  & 90.0  & 0.15  & 0.05  & 12.67  & 30    \\
        &      &       &       &       &       &       &       &        &       \\ \hline
\label{plato_table}
\end{tabular}
\end{table}

All the EBs light curves used in this work are already published and have the orbital period well established. We correctly retrieve all those measured periods. For EPIC 201408204 rotational period, we have measured as published by \citet{k22018}. Table \ref{alltable} presents all results for the orbital and rotational period for the systems observed with CoRoT, Kepler, Kepler K2 and TESS missions.

\begin{table}[h]
\caption{All retrieved quantities from sample EBs. $P_{orb}^A$ and $P_{rot}^A$ are the orbital and rotational period found in the literature identified as followed: a = \citet{corot2012}, b = \citet{kepler2011}, c = \citet{k22018}, d = \citet{tess2019}. $P_{orb}^B$ and $P_{rot}^B$ are the same quantities from this work}
\begin{tabular}{cccccccc}
ID & Name &  \begin{tabular}[c]{@{}c@{}}$P_{orb}^A$\\ {[}days{]}\end{tabular} & \begin{tabular}[c]{@{}c@{}}$P_{orb}^B$\\ {[}days{]}\end{tabular} & \begin{tabular}[c]{@{}c@{}}$P_{rot}^A$\\ {[}days{]}\end{tabular} & \begin{tabular}[c]{@{}c@{}}$P_{rot}^B$\\ {[}days{]}\end{tabular} \\ \hline
    &   &   &   &   &   \\
CoRoT 102681270    & -- & 3.0$^a$  & 6.0  & -- & 7.0  \\
    &   &   &   &   &   &   &   \\
KIC 6965293   & KOI-6800 & 5.0$^b$  & 5.0  & -- &  5.3 \\
    &   &   &   &   &   &   &   \\
EPIC 201408204    & TYC 4930-128-1 & 8.5$^c$  & 8.5 & 7.4$^c$  & 7.4  \\
    &   &   &   &   &  \\
TIC 349480507    & CD-64 270 & 1.6$^d$  & 3.1  & -- & 1.6  \\
    &   &   &   &   &   \\ \hline
\end{tabular}
\label{alltable}
\end{table}

\section{Summary and conclusion}
\label{conclusion}

We \textbf{presented} an optimized cutting off transit algorithm to study stellar rotation from space missions (CoRoT, Kepler, TESS and PLATO). We also presented some experiments to show method performance. The DT algorithm was applied in well-know CoRoT, Kepler, Kepler K2, and TESS photometric observed light curves as well as in a synthetic generated light curves expected to PLATO mission \citep{rauer2014}. \textbf{We have shown that when we precisely identify the rotation modulation and eliminate it from the light-curve, we better constrain the primary and secondary transit depths and better visually analyze the aliases on the Lomb-Scargle periodogram (see the examples of CoRoT 10268127 and TIC 34948050)}. From our analysis, we find good agreement with published measurements for the orbital period of four selected systems.  We report likely rotational periods modulation for four eclipsing binary systems, where only EPIC 201408204 has been already analyzed by \citet{k22018}. Measure rotation period for eclipsing binary is a relevant task for stellar evolution studies. These stars are fundamental targets to be used as benchmark objects, and allow an unprecedented accuracy in masses and radii of about 2$\%$.  Our rotation period measurement for  EPIC 201408204 agrees with \cite{k22018} determination.  We validated our algorithm by applying a blind test to controlled synthetic light curves with known parameters.   For PLATO\textbf{,} we generated light curves with spots modulation physics as used in $ellc$ and systematic noise as described by \citet{plato2019a}. Our code is easy to \textbf{be installed and is} ready to be used in any machine with Python3.x and with standard libraries. It is \textbf{an accessible} tool to remove \textbf{binaries} transit signals and planets in low cadence observations without detrending the entire light curve.


\bibliographystyle{spbasic}

\bibliography{references}


\end{document}